\documentclass[pre,aps,floatfix,showpacs,twocolumn]{revtex4}
\usepackage{graphicx}
\usepackage{epsfig,graphics}
\usepackage{amsmath, amsthm, amssymb}

\newcommand{\rv}{\vec{r}}
\newcommand{\tv}{\vec{t}}
\newcommand{\that}{\hat{t}}
\newcommand{\ph}{\hat{p}}

\newcommand{\be}{\begin{equation}}
\newcommand{\ee}{\end{equation}}
\newcommand{\bea}{\begin{eqnarray}}
\newcommand{\eea}{\end{eqnarray}}
\newcommand{\erf}{\operatorname{erf}}

\begin{document}
\title{Phase Transitions in  Pressurised Semiflexible Polymer 
Rings}
\author{Mithun K. Mitra}
\email{mithun@imsc.res.in} 
\affiliation{The Institute of Mathematical Sciences, 
C.I.T. Campus, Taramani, Chennai 600013, India}
\author{Gautam I. Menon}
\email{menon@imsc.res.in} 
\affiliation{The Institute of Mathematical Sciences, 
C.I.T. Campus, Taramani, Chennai 600013, India}
\author{R. Rajesh}
\email{rrajesh@imsc.res.in}
\affiliation{The Institute of Mathematical Sciences, 
C.I.T. Campus, Taramani, Chennai 600013, India}
\date{\today}

\begin{abstract}

We propose and study a model for the equilibrium
statistical mechanics of a pressurised semiflexible
polymer ring.  The Hamiltonian has a term
which couples to the algebraic area of the ring and
a term which accounts for bending (semiflexibility).
The model allows for self-intersections.  Using a
combination of Monte Carlo simulations, Flory-type
scaling theory, mean-field approximations and lattice
enumeration techniques,  we obtain a phase diagram
in which collapsed and inflated phases are separated
by a continuous transition.  The scaling properties
of the averaged area as a function of the number of
units of the ring are derived.  For large pressures,
the asymptotic behaviour of the area is calculated for
both discrete and lattice versions of the model.  For small
pressures, the area is obtained through a mapping onto
the quantum mechanical problem of an electron moving
in a magnetic field.  The simulation data agree well
with the analytic and mean-field results.

\end{abstract}
\date{\today}
\maketitle

\section{\label{intro} Introduction}

Fluid vesicles obtained via the self-assembly
of amphiphilic molecules exhibit a variety of
shapes in thermal equilibrium.  Such shapes can
be understood in terms of the energy minimising
configurations of a curvature Hamiltonian, under
the constraints of fixed enclosed volume and surface
area\cite{canham,helfrich,evans}.
Shape changes arise when solutions of the
Euler-Lagrange equations representing  distinct
shapes
exchange stability. However,
the non-linearity of these equations, if no special
symmetries are assumed, necessitates purely numerical
approaches. Further, while the curvature modulus in
bilayer lipid membrane systems is often large, so that
thermal fluctuations about the minimum free energy
structure may be 
ignored, the more general problem
of understanding the thermodynamics of such shape
transitions is a formidable one\cite{seifertreview}.

The two-dimensional version of the vesicle problem
is a polymer ring of fixed contour length, whose
enclosed area $A$ is constrained through a coupling
to a pressure difference term $p$.  Leibler, Singh and
Fisher (LSF) \cite{leibler87} performed a Monte Carlo
and scaling study of two-dimensional vesicles, modelled
as closed, planar, self-avoiding tethered chains,
accounting for both pressure and bending rigidity. In
this model, the ring polymer is obtained by connecting
the centres of impenetrable particles of fixed radius
with tethers of a fixed maximum length, while enforcing
self-avoidance. LSF showed the existence
of a phase transition at $p=0$, separating a branched
polymer phase for $p < 0$ from an inflated phase for
$p>0$. At the transition point, the ring is described
by a self avoiding polygon.  Various fractal
and non-fractal shapes that arise in these models have
also been investigated \cite{fisher89,camacho90}.

Analytic studies of this class of models present
many difficulties, arising principally from the
self-avoidance constraint. Nevertheless, the
relatively simple structure of the LSF model has
stimulated a considerable body of work, largely
in exact enumeration studies of lattice versions
of the original continuum model and its variants
\cite{rensburgbook,fisher91,melou96,guttmann,
richard01,cardy01,richard02,richard03,rajesh05}. Most of
these studies have concentrated on the behaviour of
the system in the thermodynamic limit in the region
$p \leq 0$.  However, the  $p>0$ case can exhibit
interesting crossover behaviour for large but
finite systems. 

The consequences of relaxing the
self-avoidance constraint were studied in 
Refs.~\cite{rudnick91,gaspari93,marconi93}.  In the 
models studied in these papers, the ring was allowed to
intersect 
itself, with the pressure term coupled
to the algebraic area~\cite{gaspari93,rudnick91} or to
its square~\cite{marconi93}.  The particles linked to
form the 
polymer were coupled through harmonic 
springs~\cite{gaspari93,rudnick91},
thus allowing for the extensibility of the chain.
We shall refer to this model as the Extensible
Self-Intersecting Ring (ESIR).  The ESIR model 
can be solved exactly. The solution yields collapsed and
inflated phases of the ring separated by a continuous
phase transition that occurs at a critical value of
an appropriately scaled pressure\cite{gaspari93}.
However, the model has a major shortcoming in that
the inflated phase is an unphysical one in which the
ring expands to an infinite size. In a more realistic
model, such an expansion would be limited by the finite
size of individual link lengths.

The unphysical nature of the inflated phase in the ESIR 
model has been addressed in recent work\cite{haleva06},
in which particles are joined by bonds of fixed 
length, as opposed to springs.
The Hamiltonian has
a term where the pressure couples to the algebraic
area, as in the ESIR model.  
The  transition survives as a continuous phase
transition with mean-field exponents, separating
collapsed and inflated regimes of the ring.  
We shall refer to this model
as the Inextensible Self-Intersecting Ring (ISIR).

The model proposed in this paper incorporates a bending
energy into the ISIR model
along standard lines for
semiflexible 
polymers.  We retain the coupling of
the signed pressure to the algebraic area noting, as
argued in \cite{haleva06}, that this difference, while
vastly increasing the tractability of the problem,
makes little difference to computations within the
inflated phase.
\begin{figure}
\includegraphics[width=80mm]{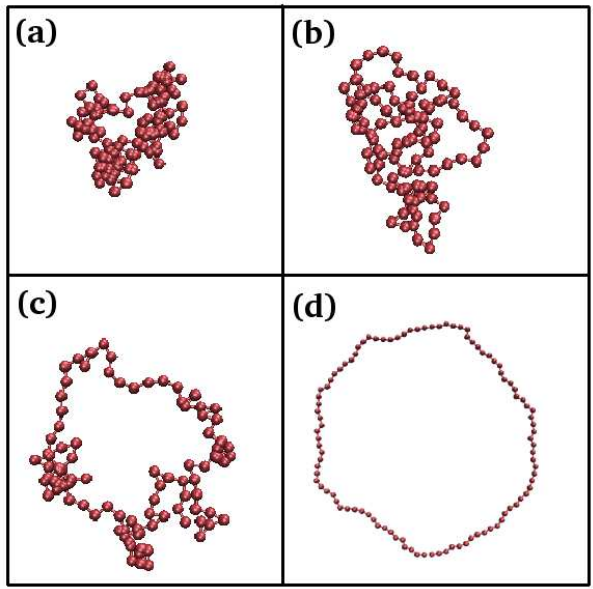}
\caption{\label{fig1}
The collapsed to inflated phase
transition as the pressure is increased. The different
panels correspond to (a) $J=0$, $p < p_c$; (b) $J=2$, $p < p_c$;
(c) $J=0$, $p = p_c$; and (d) $J=0$, $p > p_c$.}
\end{figure}

The continuum problem we address is the
following. If the polymer chain is specified by the curve 
$\rv(s)$, where $s$ is the arc-length along the curve, 
we consider the Hamiltonian 
\be
\mathcal{H} = -\frac{p}{2} \int_0^L ds \left(\rv \times \frac{d \rv}{d s} 
\right) \cdot \hat{z} + \frac{\kappa}{2} \int_0^L ds  \left|\frac{d^2 \rv}
{d s^2} \right|^2.
\ee
Here, $ds (\rv \times d \rv/d s) \cdot \hat{z}$ is the infinitesimal
(signed) area which must be integrated over the
internal variable $s$ to obtain the total algebraic
area. The quantity $L = Na$ is
the length of the polymer ring of $N$ units where $a$
is the size of the basic monomer. The continuum limit is
taken such that $N \rightarrow \infty$ and $a \rightarrow 0$,
keeping $L = Na$ fixed. The parameter $p = p_{in} -
p_{out}$ represents the pressure differential between
the inside and the outside of the ring and $\kappa$ is
the bending rigidity of the chain. We
measure energies in units of $k_B T$. The inextensibility
constraint is imposed through 
\be 
\that(s) = \left|\frac{d \rv}{d s} \right| =1, 
\ee 
where $\that(s)$ is the unit tangent
vector. When $p=0$,
this model reduces to the worm-like chain model,
with configurations  constrained by the closure
requirement. 

We will work with two discretized versions of the
above Hamiltonian.
The first has $N$ particles in the continuum
connected through fixed length links, forming a 
ring polymer whose equilibrium configurations are
constrained by the pressurisation and bending energy
terms above, but where the ring can intersect itself
at no energy cost.  We will refer to this version of
our model as the ``discrete model''.  The second is a
square lattice version of the same problem with the
particles  constrained to lie on the vertices of the
lattice. We will refer to this version as the ``lattice
model''. We discuss the differences and similarities 
between the two versions.

We use a combination of
analytic and numerical
methods to study these models:
Flory type scaling theory for the scaling of the area 
as a function of pressure, Monte Carlo simulations 
for
different pressures and bending rigidity,
mean field
approaches and exact enumeration.

In Fig.~\ref{fig1} we show typical configurations obtained
from Monte Carlo simulations of the discrete model in four
limits. These are configuration snapshots across
the collapsed to inflated phase transition, for
different values of the bending rigidity $J$ of the discrete
model ($J \propto \kappa$), as the pressure $p$
is varied. Fig.~\ref{fig1}(a) shows the collapsed phase for
the case where the bending energy is zero, while
Fig.~\ref{fig1}(b) illustrates a typical ring configuration
at an intermediate value of the bending rigidity,
but still within the collapsed regime.  In Fig.~\ref{fig1}(c),
we show a typical configuration close to the transition
between collapsed and inflated phases.  Last, Fig.~\ref{fig1}(d)
illustrates  the fully inflated ring.

\begin{figure}
\includegraphics[width=80mm]{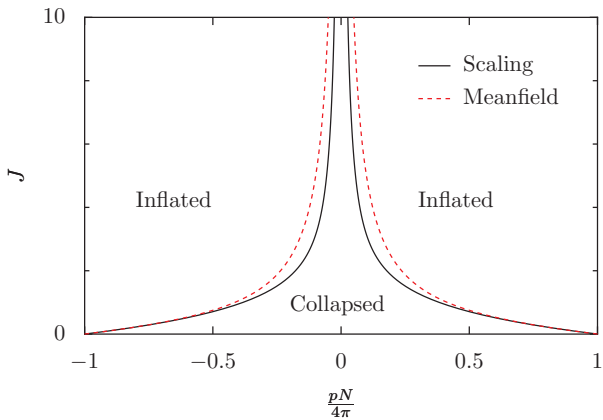}
\caption{\label{fig2}
The phase boundary between
collapsed and inflated phases for a
semi-flexible polymer ring as obtained by two different
methods, a scaling analysis based on Flory-type
arguments and mean-field theory. }
\end{figure}

We summarise our main results below. We show that there
is a continuous phase transition in the scaled pressure
$\ph~ ( N p/4\pi)$ -- bending rigidity ($J$) phase diagram, 
which separates a collapsed phase in which $\mbox{area} \propto N$,
from an inflated phase in which $\mbox{area} \propto N^2$ (see 
Fig.~\ref{fig2}). The phase boundary for the
discrete model is obtained as
$\ph_c= [I_0(J)-I_1(J)]/[I_0(J)+I_1(J)]$, where
the $I(J)$'s are modified Bessel functions. For the lattice
model, the phase boundary is obtained as
$\ph_c=e^{-J}$.

These results are obtained by  first
solving the $J=0$ case exactly and then incorporating the effects of
a nonzero $J$ through a scaling argument. For the collapsed
phase, the free energy for nonzero $J$ is calculated
by the same method. In the inflated regime,
we resort to mean field theories. We employ two types of
mean-field theories: In the first, 
the inextensibility constraint is satisfied exactly but
the closure condition is satisfied only on average. In the
second, we impose
the closure condition exactly 
but satisfy the inextensibility constraint only on 
average. The dependence of the area on $\ph$ for $\ph 
\rightarrow \infty$ is calculated. The behaviour near
the transition line is obtained through a Flory
type scaling theory.

The rest of the paper is organised as follows.
In Sec.~\ref{Model} we define our models more precisely.
Section~\ref{num_method} contains the
details of the numerical methods used, including
the Monte Carlo and exact enumeration algorithms.
In Sec.~\ref{flory}, we discuss a Flory-type
scaling theory valid for the semi-flexible case.
Section~\ref{meanfield} describes mean-field approaches
to this problem: (a) a simple density-matrix based
single-site mean-field approach which captures the
properties of the inflated phase to very high accuracy
but is inadequate for the collapsed phase and (b),
a less accurate harmonic spring mean-field theory,
which is capable of describing both collapsed and
inflated phases. 
In Sec.~\ref{pc}, we discuss the
behaviour around the critical point in greater detail.
Our Sec.~\ref{lattice} contains results for the
asymptotic behaviour of the area 
as well as a description of the appropriate scaling
function for the area in the lattice case,
as a function of $N$.
Section~\ref{conc} contains a summary and
conclusions.  In Appendix~\ref{electron} we present
a solution of the problem when $J=0$ by drawing an
analogy with the quantum mechanical problem of the
motion of an electron in a magnetic field.

\section{\label{Model} Model}

Consider a closed chain of N monomers in two dimensions. Let the 
positions of the $j^{th}$ particle be denoted by the vector $\rv_j$
and the corresponding tangent vectors by $\tv_j =
\rv_{j+1}-\rv_{j},\; j=1,2,\ldots,N.$ For a closed ring, 
$\rv_{N+1} = \rv_1$, or equivalently, $\sum_i \tv_i = 0$. 
The algebraic or signed area area $A_s$ enclosed by the ring is given by
\be
A_s = \frac{1}{2} \sum_{i=1}^{N} (\vec{r}_i \times \vec{r}_{i+1})\cdot\hat{z}
= \sum_{j=1}^{N} \sum_{k=1}^{j-1} (\tv_k \times  \tv_j) \cdot \hat{z} .
\ee
$A_s$ can be either positive or negative.

Coupling this algebraic
area to  pressure, we obtain the energy term,
\be
H_p = -p A_s. 
\ee
The bending energy cost can then be written down following standard procedures
as
\be
H_b = -  J \sum_{i=1}^{N} \that_i \cdot \that_{i+1},
\ee
where the bending rigidity $J$ of the discrete model is proportional to the 
continuum bending rigidity and $\that$ is
the unit vector in the direction of $\tv$.
The inextensibility condition is imposed through
\be
|{\vec r}_i - {\vec r}_{i-1}| = |{\vec t}_i| = a = 1.
\ee
Since the tangent vectors have unit norm, we can represent them as ${\vec
t}_i= (\cos \theta_i, \sin\theta_i)$, where $\theta \in [0, 2 \pi)$.
In terms of these variables, the partition function is
\be
\mathcal{Z} = \int \prod_{i} d\theta_i
\prod_{j=0}^{N-1} \left( \prod_{k=0}^{j-1} 
e^{\frac{p}{2} \sin(\theta_k-\theta_j)} \right) 
e^{J \cos(\theta_j-\theta_{j+1})}. 
\ee

On a square lattice, the model remains essentially the same except for
restrictions on the angles $\theta_i$. Now $\theta_i$ is allowed to take
values $0, \pi/2,\pi,3 \pi/2$ such that all the particles are on the
vertices of the square lattice.

\section{\label{num_method} Numerical method}

In this section, we describe the numerical methods used. 
For the discrete version of our model, we use Monte Carlo simulations
(described in Sec.~\ref{montecarlo}) while for the lattice
problem on the square lattice, we use an exact enumeration
scheme (described in Sec.~\ref{exact_enumeration}). The analytic 
results we obtain for our model, described in later sections, provide 
useful benchmarks for the numerical work.

\subsection{\label{montecarlo} Monte Carlo Simulations}

\begin{figure}
\includegraphics[width=80mm]{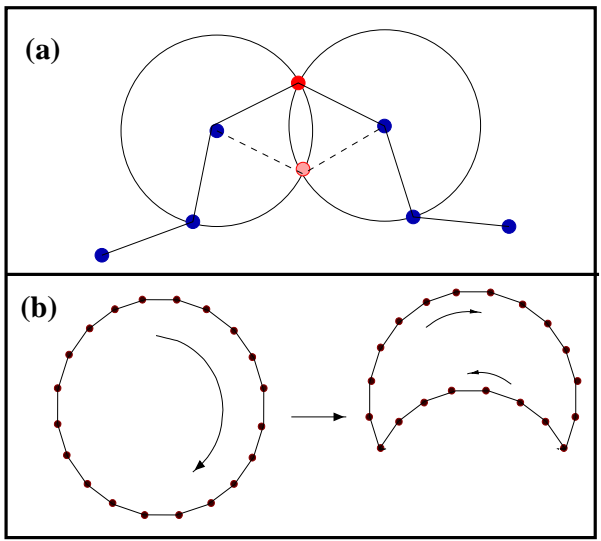}
\caption{\label{fig3} A schematic representation of the Monte Carlo moves: 
(a) single flip; and (b) global flip.}
\end{figure}
The algorithm for the Monte Carlo simulation of the
discrete model consists
of two basic moves\cite{koniaris94,rensberg90}: a single particle flip and a
global flip. In the single particle flip,
a particle is picked at random and reflected
about the straight line joining its two neighbours (see
Fig.~\ref{fig3}(a)).
The move is accepted using the standard Metropolis algorithm.
Since the energy computation involves only nearby sites,
the move is efficient and fast. In the global flip,
two particles of the ring are chosen at random
and the section of the ring between
them is reflected about the line joining the
two particles (see Fig.~\ref{fig3}(b)). 
The energy calculation now involves $O(N)$ particles and 
is thus computationally expensive. However, the global move is crucial to the 
study of the case where $J \neq 0$, since single particle moves
alone are insufficient for equilibration in this case.

In the simulations, one Monte Carlo step is defined as
one global move and $N$ single particle
moves made by selecting  at random particles to be updated. 
This step is then repeated
until the system equilibrates.
Thermodynamic
quantities are measured from averages taken over
independent configurations in equilibrium.

The initial
configuration was chosen to be a regular N-sided
polygon, but we verified that random configurations
also gave the same results.
We performed  Monte Carlo
simulations 
across a range of pressures for different values of J and system size. 
The system size varied from $N = 64$ to $N=2000$. Typically each parameter
value was run for $4 \times 10^6$ Monte Carlo steps. We waited typically for
$10^6$ steps for equilibration, averaging data over the remaining
steps using independent configurations. 

\subsection{\label{exact_enumeration} Exact enumeration}

We first describe the algorithm for the case
$J=0$.  Consider a random walk starting from
the origin and taking steps in one of the four
possible directions. For each step in the positive
(negative) $x$-direction, we assign a weight $e^{-P
y}$ ($e^{Py}$), where $y$ is the ordinate of the
walker. Multiplying these weights, it is easy to
check that the weight is $e^{PA}$
for a closed walk enclosing an area $A$.

Let $T_N(x,y)$ be the weighted sum of  all $N$-step  
walks from $(0,0)$ to $(x,y)$. 
It then obeys  the recursion relation,
\bea
T_{N+1} (x,y) &=& e^{-P y} T_N (x-1,y) + e^{P y} T_N (x+1,y) \nonumber \\
&&+ T_N (x,y-1) + T_N (x,y+1), 
\eea
with the initial condition
\be
T_0 (x,y) = \delta_{x,0} \delta_{y,0}. 
\ee
Finally, $T_N(0,0)$ gives the partition function of the ring polymer on a lattice.
 
For the semiflexible case, the  recursion relation given above
must be modified, since the ring is no longer  a simple
random walk but a walk with a one step memory. We
convert it into a Markov process as follows.
Let $T_N(x,y;x',y')$ be the sum of weights of all
walks reaching $(x,y)$ in $N$ steps but having been
at $(x',y')$ at the previous step. These $T_N$'s are
now a Markov process and depend only on $T_{N-1}$'s.
The recursion relations are then straightforward
to write down.  Rather than give all the recursion
relations, we provide a representative example
\begin{widetext}
\bea
T_{N+1} (x,y;x-1,y) &=& e^{-P y} \big[ T_{N} (x-1,y;x-2,y) + e^{2 J }  T_{N} (x-1,y;x,y)  \nonumber \\
&& +
e^J T_{N} (x-1,y,x-1,y+1) 
+ e^J T_{N} (x-1,y;x-1,y-1) \big] .
\eea
\end{widetext}
Similar recursion relations will hold for $T_{N+1} (x,y;x+1,y)$, 
$T_{N+1} (x,y;x,y-1)$ and $T_{N+1} (x,y;x,y+1)$.

The partition function for the polymer problem can be 
expressed as a sum over areas
and bends consistent with a given value of the area, i.e., 
\be
\mathcal{Z}_N = T_N (0,0) = \sum_{A,B} C_N (A,B) e^{pA + JB},
\ee
where $C_N (A,B)$ counts the number of closed paths of area $A$ in a walk of length N
which have $B$ bends.

We count up to $N=150$ for different values of $J$.
The only limiting factor in going to larger
$N$ values is computer memory. 

\section{\label{flory} Flory-type Scaling Analysis}

Flory type scaling  theory provides a useful  tool
to capture the scaling behaviour of systems whose
free energy reflects a competition between two or
more  terms.  Such a scaling theory was proposed
for the ISIR model
 in Ref.~\cite{haleva06}.  A transition  
from a collapsed to an inflated state was predicted
to occur at a critical
value of the pressure,  whose magnitude scaled with system
size as $N^{-1}$.  We show how these arguments may be
extended to the semiflexible case,
deriving expressions for the change in the critical
point and scaling as a function of the bending
rigidity.

The free energy consists of three terms describing
(i)  the  entropy of the ring, (ii)  the pressure differential and 
(iii) inextensibility of the bonds. When $J=0$, these terms were argued to
be  $R^2/N$, $-P R^2$ and $R^4/(4 N^3)$ for a ring of size $R$ \cite{haleva06}, 
where for the second term it was assumed that the area $\langle A \rangle$
scales as $R^2$.
With semiflexibility, we show that a similar scaling
form holds except for $J$ dependent prefactors. Thus,
the free energy takes the form
\bea
F &=& F_{entropic}+F_{pressure}+F_{inextensibility} , \nonumber \\
&\sim& \frac{ 4 \pi R^2}{N} \left[\alpha(J) - \ph \right]+
\frac{\beta(J) R^4}{N^3}. \label{freeen}
\eea
where we have defined $\ph = N p/4 \pi$, 
and $\alpha$ and $\beta$ depend on  $J$.

It is easily seen that a system described by
such a Flory theory undergoes a continuous
transition when the sign of the $R^2/N$ term changes sign.
This occurs at a critical scaled pressure $\ph_c(J)$ 
which varies with $J$ as
\be
\frac{\ph_c(J)}{\ph_c(0)} = \frac{\alpha(J)}{\alpha(0)}.
\ee
\begin{figure}
\includegraphics[width=80mm]{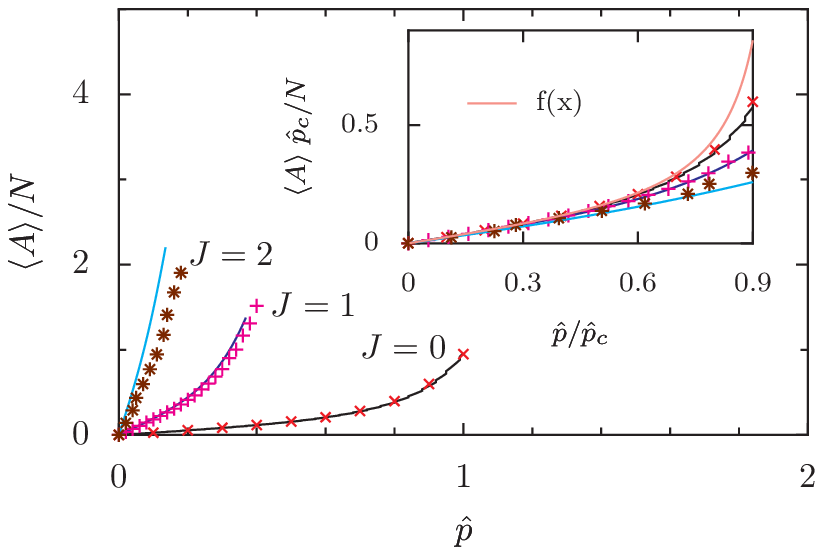}
\caption{\label{fig4} Area versus pressure curves for three
$J$ values for $\hat{p} < \hat{p}_c$. The points correspond to the
discrete case while the lines correspond to the lattice case.
The inset shows the 
collapse when the curves are scaled as in Eq.~(\ref{eq:area_scaling}).
The $f(x)$ curve in the inset represents the scaling function of Eq.~(\ref{eq:scaling_function}). 
The data is for $N=100$. }
\end{figure}

When $\ph < \ph_c(J)$, then the area follows random walk statistics with
$\langle A \rangle \sim N$. In this regime the $R^4/N^3$ term is not
important. 
For length scales much larger than the persistence length, the problem is
effectively one of a freely jointed ring, but with a suitably defined $N$.
Thus, we conclude that
\be
\langle A (J,N,\ph) \rangle = \frac{N}{\ph_c(J)} 
f\left(\frac{\ph}{\ph_c(J)}\right), \;\;\; \ph < \ph_c
\label{eq:area_scaling}
\ee
where $f(x)$ is a scaling function. The scaling function $f(x)$ and $\ph_c$
can be determined by solving the $J=0$ case (see Appendix~\ref{electron}).
This gives
\be
\ph_c=4 \pi \alpha(J),
\ee
and
\be
f(x) = \frac{1}{4 \pi x} - \frac{\cot(\pi x)}{4}.
\label{eq:scaling_function}
\ee
Numerical confirmation of Eqs.~(\ref{eq:area_scaling}) and
(\ref{eq:scaling_function}) is provided in Fig.~\ref{fig4}. 
The inset shows that the curves for different $J$ collapse onto a single
curve when scaled as in Eq.~(\ref{eq:area_scaling}). 

When $\ph=\ph_c$, the scaling is determined by the $R^4/N^3$ term. Thus,
$\langle A \rangle \sim N^{3/2}/ \sqrt{\beta(J)}$. Thus,
\be
\frac{\langle A (J) \rangle}{\langle A (0) \rangle} = \sqrt{\frac{\beta(0)}{
\beta(J)}} . \label{betaJeqn}
\ee

\begin{figure}
\includegraphics[width=80mm]{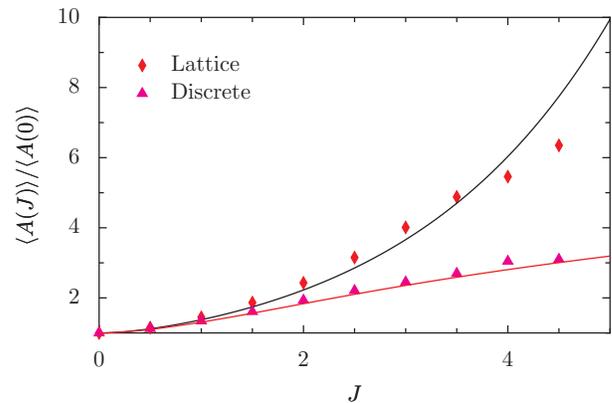}
\caption{\label{fig5} Comparison of the area ratio $\langle A (J)
\rangle/\langle A (0) \rangle$ at the critical point with the scaling prediction
(see Eqns.~(\ref{betaJeqn})) for the 
lattice (Eq.~(\ref{betalattice})) and discrete (Eq.~(\ref{betadisc})) models. 
The scaling prediction is satisfactory for small $J$ but deviates away as $J$
increases.}
\end{figure}

To test this relation, we compare
the Flory prediction with the enumeration results 
for the area in the 
lattice model. As can be seen from Fig.~\ref{fig5}, there is good agreement 
for small values of $J$ but the data starts to deviate away  
from the predicted curve as $J$ increases.

When  $\ph > \ph_c(J)$, the ring is in an inflated state, with the 
area $\langle A \rangle \sim N^2$. To obtain an accurate description of this
regime, we would need to keep higher order terms such as $R^6/N^5$ and so
on. One thus expects that the lattice and the discrete problems
should differ considerably in this regime.

We now derive expressions for $\alpha(J)$ and $\beta(J)$ 
in both the discrete and lattice cases. This is done by considering
a semiflexible chain subjected to an external force. We obtain a perturbative
solution for the partition function in the limit of small forces. From the
partition function, we obtain the free energy of the ring. By comparing
this with the form of Eq.~(\ref{freeen}),  
the values of $\alpha(J)$ and $\beta(J)$ can be obtained.

\subsection{Discrete Case}
Consider a semiflexible chain of N monomers. When the  
chain is pulled by a force $\vec{f}$, the partition function is given by
\be
Z(J,\vec{f},N) =  \int \prod_{j=1}^N d\hat{t}_j \;e^{J \hat{t}_j \cdot 
\hat{t}_{j+1}} \; e^{\vec{f} \cdot \hat{t}_j} .
\label{Zjf}
\ee
We work in the limit of small forces, treating the $J$ term 
exactly. We consider the $f$ term as a perturbation on the zeroth order 
partition function [ $f=0$ in Eq.~(\ref{Zjf})],  given by
\be
Z_0(J,N) =[2 \pi I_0(J)]^N,
\ee
where $I_0(J)$ is the modified Bessel function of the first kind of order $0$. We then expand
$\exp \left(\sum_{j=1}^{N} \vec{f}.\hat{t}_j \right)$ as a series in $f$ and average each term
with respect to the zeroth order Hamiltonian. On computing
the averages, the partition function is obtained as
\be 
\ln Z(J,f,N) =  \ln Z_0  + N b_2 f^2 + N b_4 f^4 + O(f^6),
\ee
where the coefficients $b_2$ and $b_4$ are given by
\bea
b_2 &=& \frac{I_0+I_1}{4(I_0-I_1)},\\
b_4&=& \frac{b_2^2}{4} \left[ \frac{2 I_2}{I_0-I_2} - \frac{I_0+ 3 I_1}{I_0-I_1} \right].
\eea
The $I_n$'s are modified Bessel functions of the first kind. Their
$J$ dependence has been suppressed in the equation above.

The mean end-to-end distance in the limit of small force is obtained from $R \sim \partial \ln Z /\partial f$: 
\be
\frac{R}{N}= 2 b_2 f + 4 b_4 f^3 + O(f^5).
\label{eqrn}
\ee
Solving for $f$ from Eq.~(\ref{eqrn}), we obtain
\be
f= \frac{1}{2 b_2} \frac{R}{N} - \frac{b_4}{4 b_2^4} 
\left(\frac{R}{N}\right)^3 + O\left( \left(\frac{R}{N}\right)^5 \right) .
\ee
The Flory free energy $F(R) = - \ln Z + f R$, then reduces to
\be
F(R) = -\ln Z_0 + \frac{1}{4 b_2} \frac{R^2}{N} -\frac{b_4}{16 b_2^4} \frac{R^4}{N^3} - p R^2.
\ee
Comparing with Eq.~(\ref{freeen}), the factors $\alpha(J)$ 
and $\beta(J)$ are obtained as
\bea
\alpha(J) &=& \frac{1}{4 \pi}\frac{I_0-I_1}{I_0+I_1} \stackrel{J\rightarrow
\infty}{\longrightarrow} \frac{1}{16 \pi J}, \label{alphaj}\\
\beta(J) &=& 4 \pi^2 \alpha(J)^2 \left[\frac{I_0+ 3 I_1}{I_0-I_1}  -
\frac{2 I_2}{I_0-I_2}  \right] \stackrel{J\rightarrow
\infty}{\longrightarrow} \frac{7}{64 J}. \nonumber \\ \label{betadisc}
\eea

\subsection{Lattice Case}

For a lattice polygon, where each individual step
can point only in four directions, we solve the problem 
of a semiflexible chain subject to an external force
using the exact $4 \times 4$ transfer matrix.
The transfer matrix in this case is given by
\be
T = \left( \begin{array}{llll}
					e^{J+f} & e^{f/2} & e^{-J} & e^{f/2} \\
					e^{f/2} & e^{J} & e^{-f/2} & e^{-J} \\
					e^{-J} & e^{-f/2} & e^{J-f} & e^{-f/2} \\
					e^{f/2} & e^{-J} & e^{-f/2} & e^{J} 
				 \end{array}
			\right)
\ee
We determine the largest eigenvalue up to order $f^4$, and hence calculate the 
partition function:
\bea
\ln Z (J,f,N) &=& N \left[ \ln (2 + e^{-J} + e^{J}) + \frac{e^J}{4} f^2 \right. \nonumber \\
&&+ \left. \frac{1}{192} (e^J - 3 e^{3 J}) f^4 +O(f^6)\right] .
\eea
We then follow the same procedure as for the discrete case, finding $R/N$ in terms of $f$, inverting this
equation to find $f$, and finally using this expression to compute the free energy. We thus obtain
\be
F(R)
= e^{-J} \frac{R^2}{N}  +\left[ \frac{1}{12} e^{-3 J} (3 e^{2 J} -1 ) \right]  \frac{R^4}{N^3} .
\ee
The expressions for $\alpha(J)$ and $\beta(J)$ are then
\bea
\alpha(J) &=& \frac{1}{4 \pi} e^{-J} , \\
\beta(J) &=& \frac{1}{12} e^{-3 J} (3 e^{2 J} -1 ) . \label{betalattice}
\eea

\section{\label{meanfield} Mean Field Theory}

In this section we present mean-field theories to calculate
the dependence of area on pressure and bending rigidity. In
Sec.~\ref{density}, we address the ISIR model ($J=0$). The 
mean field theory presented in \cite{haleva06} 
performs
poorly with respect to
the Monte Carlo data when $\ph > \ph_c$. Here,
we present an improved variational mean field which reproduces the
behaviour of the area above the transition very accurately.  It also
yields the correct asymptotic behaviour for the area in
the limit of high pressures. In this approach, the constraint
of fixed link length in treated exactly while the closure
constraint is satisfied in a mean field sense. However,
such a mean-field theory fails to describe the
collapsed phase, also yielding incorrect results for
the case of nonzero $J$. 

In Sec.~\ref{harmonic}, we generalise an earlier
mean field theory for the freely jointed chain to
include semi-flexibility, imposing the constraint
of fixed bond length via a Lagrange multiplier
\cite{gaspari93,haleva06}. The closure condition
is imposed exactly.  We thus derive expressions for
the average area of the ring for all pressures and
bending rigidity.

\subsection{\label{density} Density matrix mean-field for flexible polymers}
In variational 
theory, a trial density
matrix $\rho$ is chosen to approximate the actual
density matrix\cite{chaikinlubensky}. The variational
parameters are determined by 
minimising the
variational free energy $F_{\rho}$ with respect to the
parameters. The simplest mean-field 
theories
assume a trial density matrix that is a product
of independent single particle matrices, i.e,
\be 
\rho = \prod_j \rho_j, 
\ee
where $\rho_j$
is the single particle density matrix
of particle $j$. 
The variational mean-field free energy is
\be 
F_{\rho} = \langle \mathcal{H} \rangle_{\rho} + T \sum_j Tr \rho_j \ln \rho_j. 
\ee
The variational form for the density matrix should satisfy the constraint
$\rm{Tr}~\rho_j =1$.
\begin {figure}
\includegraphics[width=80mm]{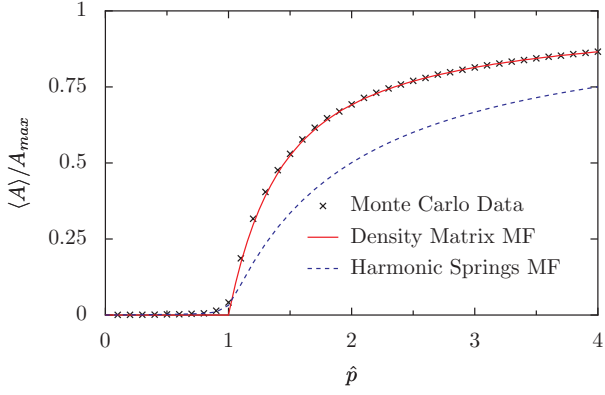}
\caption{\label{fig6} Comparison of Monte Carlo data with the two 
meanfields for the flexible ($J=0$) case.
The density matrix-based mean-field approach provides an accurate description of the 
area for $\ph> \ph_c$.}
\end {figure}

We choose the single particle density matrix based on the high pressure
limit. In this limit, the ground state of our Hamiltonian is a regular N-gon, 
where the angle of the $j^{th}$ tangent vector is $\theta_j=2 \pi j/N$. 
The single 
particle density matrix has a delta function peak at this value. At 
intermediate pressures, we therefore take the form of the density matrix 
to be a gaussian of width $\sigma$ (the variational parameter)
centered  about $\theta_j$:
\be 
\rho_j(\theta_j) = \frac{1}{\sqrt{2 \pi} 
\sigma \erf[\pi/\sqrt{2}\sigma]} \exp \left[\frac{-(\theta_j-
\frac{2 \pi j}{N})^2}{2
\sigma^2}\right],
\ee
where the normalisation ensures that 
$\rm{Tr}~\rho_j =1$ and $\erf(x)$ is the error function defined as
\be
\erf(x) = \frac{2}{\surd\pi}\int_0^x e^{-t^2}\,dt.
\ee

Using this form of the density matrix, we obtain
\bea
\lefteqn{\frac{F_{\rho}}{N} = 
-\frac{p}{4} \cot\left(\frac{\pi}{N}\right) K(\sigma)^2 + 
 J \cos\left(\frac{2 \pi}{N}\right) K(\sigma)^2}  \nonumber \\
&& - \frac{1}{2} + \frac{\sqrt{\pi}\exp(\pi^2/(2 \sigma^2))
}{\sqrt{2}\sigma \erf[\pi/\sqrt{2} \sigma]} - \ln
\left(\!\!\! \sqrt{2 \pi} \sigma \erf[\frac{\pi}{\sqrt{2} \sigma}]\right),
\quad \label{fvar}
\eea
where
\be
K(\sigma) = \frac{ \erf[(\pi - \imath \sigma^2)/\sqrt{2}
\sigma] + \erf[(\pi + \imath \sigma^2)/\sqrt{2} \sigma]
}{2 \erf[\pi/\sqrt{2}\sigma] e^{\sigma^2/2}}.
\ee
When $N\gg 1$, the pressure and bending terms in Eq.~(\ref{fvar})
can be combined, and the problem is equivalent to
one of a flexible polymer ($J=0$) with an effective 
pressure  $\ph_{\mathrm{\rm{eff}}} = \ph+J$.

The variational parameter $\sigma$ is chosen to be the $\sigma^*$ that
minimises $F_{\rho}$ in Eq.~(\ref{fvar}).
This is done numerically. The average area, equal to 
$-\partial F_{\rho} / \partial p$, is then given by
\be 
\langle A \rangle = \frac{N}{4} \cot\left(\frac{\pi}{N}\right) K^2(\sigma^*)
\stackrel{N\rightarrow \infty}{\longrightarrow}
\frac{N^2}{4 \pi} K^2(\sigma^*).
\ee 
\begin {figure}
\includegraphics[width=80mm]{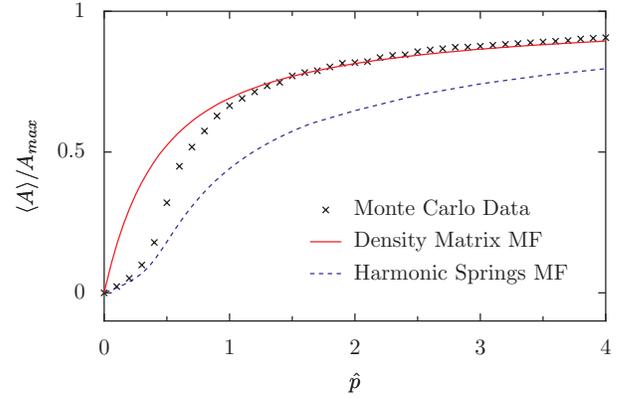}
\caption{\label{fig7} Comparison of Monte Carlo data with the two 
mean-field approaches for the case  $J=1$.}
\end {figure}

We now derive 
the asymptotic behaviour of area 
in the limit of high pressures. We work in the limit when $N$ is large.
For large pressures, we expect that $\sigma^*$ tends to zero. In this limit
\be
K(\sigma) \approx e^{-\sigma^2/2}, \quad \sigma \rightarrow 0.
\ee
and the variational free energy is then given by 
\be 
F_{\rho}(\sigma) = N\left[-(\hat{p}+J)e^{-\sigma^2} - \ln (\sqrt{2 \pi} \sigma) -\frac{1}{2}
\right], 
\ee
where $\hat{p} = N p/(4 \pi)$.
Solving $d F_{\rho}/d \sigma^*=0$, it is straightforward to obtain
\be
\sigma^* = \frac{1}{\sqrt{2 \hat{p}}} + \frac{1-2 J}{4 \sqrt{2} \hat{p}^{3/2}}, \quad \hat{p} \rightarrow \infty.
\ee
The area then reduces to
\be 
\frac{\langle A \rangle}{N^2/4 \pi} \rightarrow 
1 - \frac{1}{2 \hat{p}} +\frac{4 J -1} {8 \hat{p}^2} , \quad \hat{p} \rightarrow \infty. \label{asymptoteJ0mf}
\ee

For flexible polymers ( $J=0$), this mean-field theory reproduces the 
$\hat{p} > \hat{p}_c$ behaviour very accurately.
It also obtains the correct asymptotic behaviour.
In Fig.~\ref{fig6}, we compare the Monte Carlo data for $J=0$ 
with the results of 
the above  mean field theory and contrast it with  
the meanfield theory of Ref.~\cite{haleva06}. 

The density matrix mean-field however, fails to
correctly obtain the behaviour for non-zero values of
the bending rigidity. 
It predicts a first order transition
for $J\geq 1$, in disagreement with results from scaling theory.
We compare the results of this mean field with the Monte
Carlo data in Fig.~\ref{fig7} for a system with $J=1$. This mean-field
approach then predicts a transition at $\ph = 0$. The discrepancy
between the two curves increases for larger values of $J$.

We now describe an alternative mean-field approach to this problem
which extends the harmonic spring-based mean field theory
of Ref.~\cite{haleva06} to non-zero values of $J$.

\subsection{\label{harmonic} Harmonic spring mean-field for semiflexible polymers}

We follow the approach of
Refs.~\cite{gaspari93,haleva06} wherein the
rigid links between particles are replaced by extensible springs.
The spring constant
$\lambda$ of the springs is identified with a Lagrange multiplier,
chosen so that the mean length of a spring equals unity.

Consider a partition function for N particles given by,
\be
\mathcal{Z}\! =\! \!\! \int\!\! d\tv_j \exp\!\! \left[ \frac{p}{2} \sum_{k < j} \tv_k \times \tv_j
 + J \sum_j \hat{t}_j \cdot \hat{t}_{j+1} \!-\! \lambda \sum_j \tv_j^2 \right].
\ee
Note that while pressure couples to $\tv$, the bending rigidity
couples to the unit vectors $\that$. We make the approximation of
replacing $\that$ by $\tv$. This makes the problem analytically
tractable. 

Expanding the tangent vectors in Fourier space as,
\bea
\hat{t}_{j}^{x} &=& \sqrt{\frac{2}{N}} \sum_k [ A_k \cos(jk) + B_k \sin(jk) ], \nonumber \\
\hat{t}_{j}^{y} &=& \sqrt{\frac{2}{N}} \sum_k [ A_k^{'} \cos(jk) + B_k^{'}
\sin(jk) ],
\eea
where $ k = 2 \pi l/N$, $l=1,2,\cdots,N$.
The partition function then reduces to
\bea
\lefteqn{\mathcal{Z} = 
\prod_k \int dA_k dA_k^{'} dB_k dB_k^{'}} \nonumber \\
&&e^{-(\lambda-J \cos k)( A_k^2 + B_k^2 + {A_k^{'}}^2 + {B_k^{'}}^2 )}
e^{\frac{p}{k} (B_k A_k^{'} - A_k B_k^{'})} .
\eea
By completing the squares, this integral can be written 
as a gaussian integral and hence can be calculated exactly. This gives
\be
\mathcal{Z} = \prod_k \frac{1}{\lambda-J \cos k} \times
\left[ 1-\frac{p^2}{4 k^2 (\lambda-J \cos k)^2} \right]^2 .
\ee
The parameter $\lambda^*$ is determined by equating the mean square link
length to one, i.e
\be
-\frac{1}{N} \frac{\partial \ln \mathcal{Z}}{\partial \lambda}=1 .
\ee
This gives
\be
N = \sum_{l=1}^N \frac{1}{\lambda^*-J \cos (\frac{2 \pi l}{N})} 
\left[1 \!+\! \frac{2 \hat{p}^2}{l^2 [\lambda^*-J
\cos (\frac{2 \pi l}{N})]^2 \!-\! \hat{p}^2} \right],
\label{lambdastar}
\ee
where $\ph = pN/4 \pi$. 

When $J=0$, the first factor in Eq.~(\ref{lambdastar})
becomes independent of $l$, and then the resultant
expression can be evaluated exactly. Hence, an analytic
expression for $\lambda^*$ can be obtained in this case
\cite{haleva06}. For $J \neq 0$, this is no
longer possible, and for finite system sizes
the resultant equation must be solved numerically. When $N\gg 1$,
it is still possible
to extract the behaviour of the system analytically.

We now determine the phase boundary from
Eq.~(\ref{lambdastar}).  We will consider the limit
$N\gg 1$. First, note that $\lambda^*-J \cos (2 \pi
l/N) \neq 0$ for all $l$. For positive $\lambda^*$,
this gives the condition that $\lambda^* > J$. Second,
consider the term in the denominator for $l=1$. It
is $(\lambda^*-J)^2-\hat{p}^2$. If we assume that
$\lambda^*$ is continuous in $\hat{p}$, we have
the second constraint that $\lambda^* >J+\hat{p}$.

Setting $x=\frac{l}{N}$ and converting the first sum in
Eq.~(\ref{lambdastar}) to an integral, the equation for $\lambda^*$ reduces
to 
\bea
\lefteqn{1 =  \frac{1}{\sqrt{\lambda^{*2}-J^2}} - \frac{1}{N(\lambda^*-J)}  }\nonumber \\
&&+  \frac{2}{N(\lambda^*-J)}
\sum_{k=1}^{\infty} \left(
\frac{\ph}{\lambda^*-J}\right)^{2k} \!\! \frac{1}{2k-1} + \mathcal{O}(\frac{1}{N^2}).  \quad
 \label{lambdastarint}
\eea
The sum in Eq.~(\ref{lambdastarint}) is convergent if the ratio $\ph/(\lambda^*-J) < 1$. In this case,
we keep only the first term on the right hand side of Eq.~(\ref{lambdastarint}). This gives,
\be
\lambda^* = \sqrt{1+J^2}, \;\;\; \mathrm{for}\; \ph < \ph_c .
\ee

The critical pressure is obtained when the ratio $\ph/(\lambda^*-J)$ becomes equal to $1$, i.e.
\be
\ph_c(J) = \lambda^* - J = \sqrt{1+J^2}-J .
\ee
For large values of $J$, this goes as $\ph_c(J) \sim 1/2 J$, which differs by a factor
of $2$ from the answer obtained by scaling arguments [see Eq.~(\ref{alphaj})].

We shall now estimate $\lambda^*$ in the different scaling regimes. 
We assume that $\lambda^*$ is a non-decreasing function of $\ph$ (as in $J=0$).
Then, since we have the constraint of $\lambda^* > \ph + J$, the ratio 
$\ph/(\lambda^*-J)$ must continue to remain at $1$ for $\ph > \ph_c$.
Thus, above the critical point, we obtain
\be
\lambda^* = \ph +J, \;\;\; \mathrm{for}\; \ph > \ph_c . \label{lambdap>pc}
\ee

However, a simple substitution of Eq.~(\ref{lambdap>pc}) in Eq.~(\ref{lambdastar}) for
$\ph > \ph_c$ does not satisfy Eq.~(\ref{lambdastar}). We therefore need to calculate 
the correction term arising from large but finite $N$. We start by considering 
Eq.~(\ref{lambdastar}). The first term can be summed exactly, giving
\bea
\lefteqn{N \left( 1 - \frac{1}{\sqrt{\lambda^2-J^2}} \right)}
 \nonumber \\
&&= \sum_{l=1}^N \frac{1}{\lambda-J \cos(\frac{2 \pi l}{N})} \frac{2 \ph^2}
{l^2 (\lambda - J \cos(\frac{2 \pi l}{N}))^2-\ph^2}.
 \label{lambda}
\eea
We calculate the finite size corrections to $\lambda^*$ as follows.
Let
\be
\lambda^*_{\ph \geq \ph_c} = \ph +J-\delta .
\ee
When $\delta \rightarrow 0$, the main contribution to the left hand side of Eq.~(\ref{lambda})
comes from the $l=1$ term. The contribution from other $l$ is convergent as $\delta \rightarrow 0$.
Expanding the right hand side as a series in $\delta$, we obtain
\be
-\frac{1}{\delta} = N\left[ 1-\frac{1}{\sqrt{\ph^2 + 2 \ph J}} -\frac{ \delta (\ph+J)}
{( \ph^2 + 2 \ph J)^{3/2}} \right] .
\label{lambdapc}
\ee
The $\delta$ independent term in the right hand side of Eq.~(\ref{lambdapc}) is nonzero
for $\ph > \ph_c$ and is equal to zero for $\ph=\ph_c$. Thus, when $\ph  
> \ph_c$, we keep
only the first term in the right side, while at $\ph=\ph_c$, we need to keep the second term too.
Solving for $\delta$, we obtain,
\be
\label{eq:delta}
\delta  =  
\begin{cases} 
\frac{1}{\sqrt{N}} \frac{1}{(1+J^2)^{1/4}} ,& \text{ $\ph=\ph_c$,}\\
\frac{1}{N}  \frac{\sqrt{\ph^2+2\ph J}} {\sqrt{\ph^2+2\ph J}-1} , & \text{ $\ph>\ph_c$.}
\end{cases}
\ee

We are now in a position to calculate the mean area $\langle A \rangle$ 
from $\frac{\partial \ln \mathcal{Z}}{\partial p}$. 
This gives,
\be
\langle A \rangle = \frac{N \hat{p}}{2 \pi}
\sum_{l=1}^N \frac{1}{l^2 (\lambda^*-J \cos (\frac{2 \pi l}{N}))^2-\hat{p}^2}. \label{areaHMF}
\ee
The numerical values obtained for $\lambda$ are then substituted in this equation to get the 
corresponding value of the area. We can, however, analytically
determine the scaling behaviour of the area 
in the limit of large system sizes from the values of $\lambda$ calculated above.

For $\ph < \ph_c$,  we have,
\be
\langle A \rangle \simeq \frac{N \hat{p}}{2 \pi} \sum_{l=1}^N \frac{1}{l^2(\sqrt{1+J^2}-J \cos(2 \pi l/N))^2-\ph^2}
\ee
At the critical point, we obtain, from Eqns.~(\ref{lambdapc}) and
(\ref{areaHMF}),
\be
\langle A \rangle = N^{3/2} \frac{(1+J^2)^{1/4}}{4 \pi}, \quad \ph = \ph_c.
\ee
Similarly, for pressures greater than the critical pressure, we obtain, from
Eqns.~(\ref{lambdap>pc}) and (\ref{areaHMF}),
\be
\langle A \rangle = \frac{N^2}{4 \pi} \left[1-\frac{1}{\sqrt{\ph^2+2\ph
J}}\right]  \stackrel{\ph \rightarrow \infty}{\longrightarrow} \frac{1}{\ph} -
\frac{ J}{2 \ph^2}, 
~ \ph > \ph_c.
\ee

This mean field theory reproduces the qualitative behaviour of the
simulation data correctly. It
predicts a continuous transition for all $J$, unlike the density matrix 
field theory. However, there is a quantitative disagreement with the
data.
This can be seen by comparing the results of this
mean-field theory with the simulation data in both the 
flexible (Fig.~\ref{fig6}) and semi-flexible (Fig.~\ref{fig7}) polymer 
cases. 

\section{\label{pc} Scaling and Critical Exponents}

\begin{figure}
\includegraphics[width=80mm]{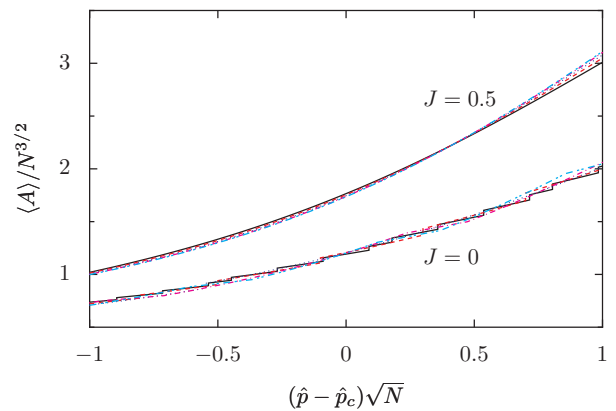}
\caption{\label{fig8} 
Area collapse for flexible and semiflexible polymers
around the critical point. 
This verifies Eq.~(\ref{Ascaling}). The data is for $N=80,100,120,140,150$ for
the lattice problem. 
}
\end{figure}
The order parameter that describes the collapsed to
inflated phase transition is the ratio of the area
to the maximum area. When $N\gg 1$, the ratio is
zero below the transition
and non-zero above it. The behaviour near the 
transition line can be described by the scaling form
\be
\frac{\langle A \rangle}{A_{max}} \simeq N^{-\phi \beta} 
g\left[ (\ph-\ph_c) N^\phi \right], \label{Ascaling}
\ee
where $\phi, \beta$ are exponents and $g(x)$ is a scaling function.
When $x\rightarrow 0$, then $g(x) \rightarrow \rm{constant}$. 
When $x\rightarrow \infty$, then $g(x) \sim x^\beta$.  
When $x \rightarrow  -\infty$, then $g(x) \sim 1/x$
[see Eqs.~(\ref{eq:area_scaling}) and (\ref{eq:scaling_function})]. This
immediately implies that
\be
\phi (1+\beta)=1.
\label{exponent_relation}
\ee

To obtain the one independent exponent, we resort to
the scaling theory (see Sec.~\ref{flory}). At $\ph_c$,
$\langle A \rangle/A_{\rm{max} }\sim 1/\sqrt{N}$.
At the critical point, the area
scales as $N^{3/2}$.  Combing with Eq.~(\ref{exponent_relation}),
we obtain $\phi=1/2$ and $\beta=1$. These exponents are independent of
$J$.

In  Fig.~\ref{fig8}, we show scaling plots
when area is scaled as in Eq.~(\ref{Ascaling}) with $\phi$
and $\beta$ as above for the cases $J=0$ and $J=0.5$. 
The excellent collapse shows that the Flory type
scaling theory gives the correct
exponents.

We now look at the fluctuations. Consider the compressibility $\chi$ defined
as
\be
\chi=\frac{1}{A_{\rm{max}}} \frac{\partial \langle A \rangle }{\partial p}.
\ee
When $\ph < \ph_c$, $\chi$ can be calculated from
Eqs.~(\ref{eq:area_scaling}) and (\ref{eq:scaling_function}) to be
\be
\chi = - \frac{1}{\ph^2} + \frac{\pi^2}{\ph_c^2 \sin^2(\pi \ph /\ph_c)},
~~\ph<\ph_c. \label{chiplesspc}
\ee
Thus, $\chi$ diverges as $(\ph_c-\ph)^{- 2}$ below the transition point. The
behaviour near the transition point is described by the scaling form
\be
\chi
\simeq N^{\phi \gamma} 
h\left[ (\ph-\ph_c) N^\phi \right], \label{chiscaling}
\ee
where $h(x)$ is a scaling function and $\phi=1/2$. 
When $x\rightarrow 0$, then $h(x)
\rightarrow \rm{constant}$. When $|x| \gg 1$, then $h(x) \sim x^{-\gamma}$. 
Comparison with Eq.~(\ref{chiplesspc}) gives $\gamma=2$.

\begin{figure}
\includegraphics[width=80mm]{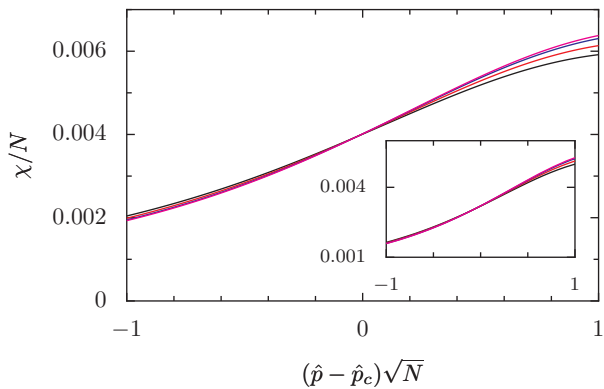}
\caption{\label{fig9} The scaling plots for compressibility $\chi$ when
scaled as in Eq.~(\ref{chiscaling}). The data is for the lattice model
with $J=0.5$ and $J=0$
(inset). The system sizes are 
$N=80,100,120,140,150$.}
\end{figure}

In Fig.~\ref{fig9}, we plot the compressibility scaled as in
Eq.~(\ref{chiscaling}) for two different values of $J$. A good collapse is
obtained again showing that the Flory type scaling theory gives the correct
exponents. Similar, but noisier data can be obtained for the discrete model.
We thus conclude that the introduction of semiflexibility does not affect 
any of the
exponents describing the transition. 

\section{\label{lattice} The lattice problem}

In this section, we present some additional enumeration results for the
lattice problem. Consider the scaling theory presented in Sec.~\ref{flory}.
The inextensibility of the polymer was captured by
the $R^4/N^3$ term for a polymer of extent $R$. This was obtained from a
calculation based on the extension of a polymer under a force. Here we
present numerical evidence supporting this. 

Let $P_N(A)$ be the probability
(at $P=0$) that a walk of length
$N$ encloses an area $A$. In Appendix~\ref{electron}, we
show that [see Eq.~(\ref{eq:cn_appendix})]
\be
P_N(A) = \frac{1}{N} I\left(\frac{A}{N} \right), ~A,N\rightarrow \infty,
~\frac{A}{N} ~\rm{fixed}.
\label{cn_scaling}
\ee
where the scaling function $ I(x) $ is given by
\be
I(x) = \pi ~\mathrm{sech}^2(2 \pi x ).
\ee

\begin {figure}
\includegraphics[width=80mm]{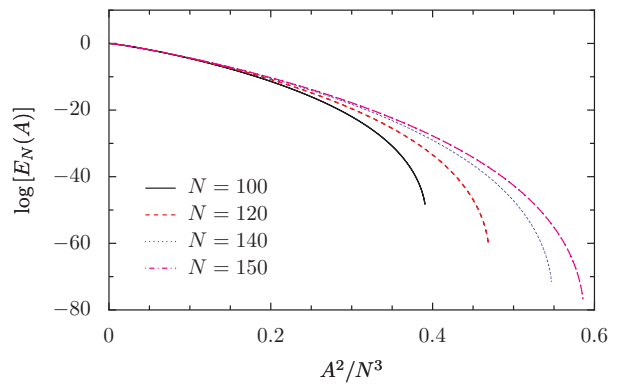}
\caption{\label{fig10} Collapse of the $E_N(A)$ for different values of 
$N$ when plotted against $A^2/N^3$. The data is for the lattice model with
$J=0$.  }
\end {figure}

We consider the corrections to the scaling form in Eq.~(\ref{cn_scaling}).
Let
\be
E_N(A) = \frac{N P_N(A)}{ I(A/N)}.
\ee
Scaling theory predicts that $E_N(A)$ should be a function of one
variable $A^2/N^3$. This is verified in Fig.~\ref{fig10} where
$\ln E_N(A)$ is plotted against $A^2/N^3$ for a range of
system sizes. 
\begin {figure}
\includegraphics[width=80mm]{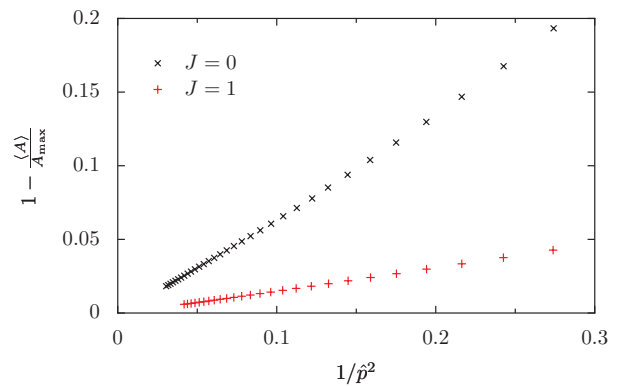}
\caption{\label{fig11} The asymptotic behaviour of area in the limit
of large $\ph$ as computed for the lattice model. 
The curves are straight lines when plotted against
$1/\ph^2$.}
\end {figure}

We also study the behaviour of area when $\ph$ is very large. When $\ph \gg
1$, the behaviour is seen to differ from the discrete version of the
problem. It can be shown to be\cite{mithununpublished}
\be
1- \frac{\langle A \rangle }{A_{\rm{max}}}  \sim \frac{1}{\ph^2}, \quad \ph
\rightarrow \infty. 
\label{lattice_asymptotic}
\ee
This should be contrasted with the discrete case which varied as $1/\ph$. In
Fig.~\ref{fig11}, we show numerical confirmation of the prediction of
Eq.~(\ref{lattice_asymptotic}).

\section{\label{conc} Conclusions}

In this paper, we have proposed and studied lattice and
discrete models for self-intersecting pressurised
semi-flexible polymers. Our work generalises results of
Ref.~\cite{haleva06} to include a bending rigidity. A
simple variational mean-field approach provides very
accurate fits to the  Monte Carlo data for this problem
in the absence of semi-flexibility. The  mean-field
approach for $J=0$ \cite{rudnick91,gaspari93,haleva06}
was generalised to the semiflexible case. The phase
boundary between collapsed and inflated phases as well as
expressions for the area as a function of $p$
and  $J$ in the different phases were obtained 
analytically. 

We have shown that the essence of the physics is
captured through 
simple Flory approximations.
The scaling predictions of the Flory theory were 
verified numerically for both the lattice and discrete cases.

We have also investigated the behaviour of the system in
the extreme limits of a fully pressurised polymer
ring and a collapsed configuration. For the fully pressurised
ring, we deduce the leading order asymptotic behaviour of the
area in both the discrete and lattice cases. The collapsed
phase was studied by mapping this problem onto a quantum mechanical 
problem of an electron confined to two dimensions and placed in a
transverse magnetic field.  The analytic results thus
obtained fit the data accurately.

The usefulness of these results for more realistic
systems lies in the fact that both the restriction
to the signed area as well as allowing for
self-intersections at no energy cost are irrelevant
in the large $p$ limit.
The results obtained at
large $p$ should therefore apply both qualitatively
and quantitatively to the more realistic case of a 
pressurised self-avoiding polymer, where the pressure
term couples to the true physical area and not to
the signed area. This is the LSF model
\cite{leibler87}. The approach presented here is thus
also useful in understanding the behaviour of a larger
class of models, some of which are more physical in 
character, but which lack the analytic tractability
of the model proposed and studied here.

\begin{acknowledgments}
This work was partially supported by grant
3504-2 of the Indo-French Centre for the Promotion of
Advanced Research and the DST, India (GIM).
\end{acknowledgments}

\appendix

\section{\label{electron} Analytic answer in the small pressure regime}

The problem of self-intersecting polymers in two dimensions with no
bending rigidity ($J=0$) is analogous to
the quantum mechanical problem of an electron moving in a magnetic field 
applied transverse to the plane of motion. 
Using this analogy, we obtain analytic expressions for the partition 
function $\mathcal{Z}$ and $C_N(A)$, the number of closed walks of area $A$.

When an electron 
goes around a magnetic field, it picks up a phase factor 
proportional to the flux enclosed by the path. This flux  is proportional 
to the product of the strength of the magnetic field times the algebraic area
enclosed by the loop. The propagator then is the sum over all such loops. 
This suggests that the partition function for the polymer problem 
can be obtained from the quantum
mechanical propagator for the electron problem provided the constants are
appropriately mapped.

For an electron of charge $e$ and mass $m$ in a constant external 
magnetic field $B$, in the z direction, in the case when the electron 
returns to the origin, the kernel can be written as \cite{feynmanbook},
\be 
K(0,0;t,0) = \left( \frac{m}{2 \pi i \hbar t} \right) \left( \frac{\omega
t/2}{\sin \omega t/2} \right), 
\ee
where $\omega = e B/m c$.
It picks up a flux $\Phi$  given by
\be 
\Phi = \frac{e B A}{\hbar c}. 
\ee

The motion of a quantum mechanical particle is governed by the two-dimensional
Schroedinger equation,
\be
\imath \hbar \frac{\partial \psi}{\partial t} = \frac{- \hbar^2}{2 m} 
\left( \frac{\partial^2 \psi}{\partial x^2} + \frac{\partial^2 \psi}{\partial y^2} \right).
\ee
The classical diffusion equation for a particle in two-dimensions is
\be
\frac{\partial P}{\partial t} = \frac{1}{4} \left( 
\frac{\partial^2 P}{\partial x^2} + \frac{\partial^2 P}{\partial y^2} \right),
\ee 
where 
$P(x,y)$ is the probability of finding the particle at $(x,y)$.
Thus to map the results of the quantum problem onto the polymer problem, we need to map $t
\rightarrow -it$  and also identify
\be
\frac{\hbar}{2 m} = \frac{1}{4} .
\ee
Also, to match the diffusion constant, we need,
\be
\frac{\imath e B }{\hbar c} = p .
\ee
Substituting in the propagator and equating it to the partition function,
we obtain
\be
\mathcal{Z} = \frac{4^N}{4 \pi} \frac{p}{\sin \frac{p N}{4}}.
\ee
where we have introduced the appropriate normalisation factors.
$C_N (A)$ is now obtained from the partition function by performing the inverse 
Laplace Transform with respect to $p$.  This gives
\be
C_N(A) = \frac{ 4^{N+1}}{ 2 N^2} \mathrm{sech}^2 \frac{2 \pi A}{N}.
\label{eq:cn_appendix}
\ee
and 
\be
\langle A \rangle = \frac{1}{p} - \frac{N}{4} \cot \left(\frac{N p}{4} \right).
\ee

The free energy will have a singularity at $p=4\pi/N$. Below this $p$, the
expressions are valid for both the discrete case and the lattice.
Exactly the same expression has been obtained by using the harmonic spring
approximation \cite{rudnick91}.
The expression for area matches 
both the simulation
and lattice data quite closely for low pressures, as can
be seen from Fig.~\ref{fig4}.

Moreover, if we recall the Flory prediction that by rescaling 
area and pressure by $\ph_c(J)$, we can obtain the results for
non-zero values of the bending rigidity from the answer of the
problem with $J=0$, we see that the above analysis also predicts 
the area expression for nonzero values of $J$.

\end{document}